\documentstyle[epsf,11pt]{article}
\def\ll{\label}
\def\re{\ref}
\def\c{\cite}

\def\r1{(\ref{$1})}

\def\ba{\begin{array}{c}}

\def\ea{\end{array}}

\def\ov{\over}
\def\ha{{1\over 2}}

\def\l{\left}
\def\l({\left(}
\def\r){\right)}
\def\r{\right}

\def\al{\alpha}

\def\be{\begin{equation}}
\def\bc{\begin{center}}
\def\ec{\end{center}}
\def\bit{\begin{itemize}}
\def\eit{\end{itemize}}
\def\ee{\end{equation}}
\def\ed{\end{document}}
\def\bea{\begin{eqnarray}}
\def\eea{\end{eqnarray}}
\def\efr{\end{flushright}}

\begin{document}
\title{New series of integrable vertex models
through a unifying approach
}

\author{
Anjan Kundu \footnote {email: anjan@tnp.saha.ernet.in} \\  
  Saha Institute of Nuclear Physics,  
 Theory Group \\
 1/AF Bidhan Nagar, Calcutta 700 064, India.
 }
\maketitle
\vskip 1 cm

\begin{abstract} 
Applying a  unifying Lax operator approach   to statistical systems a new
 class of  integrable vertex models based on quantum algebras is proposed,
 which exhibits a rich variety for generic q, q roots of unity and $q \to 1$.
 Exact solutions are  formulated through algebraic Bethe ansatz 
  and a novel possibility  of hybrid vertex models is introduced.
    
-----
\medskip
\end{abstract}

 PACS numbers 02.30.Ik,
  02.20.Uw,
05.20.-y
05.50+q,
03.65.Fd

\smallskip

  
\noindent {\it Introduction}:

$D$-dimensional quantum systems are  known to be related to $(1+D)$-dimensional
classical statistical models, which is true   naturally also
 in $D=1$, where  one  finds an exclusive  class of
   models, known as integrable systems, allowing exact solutions.
 Celebrated example of such relation is that between the $XYZ$ quantum
spin-$\ha$ chain and the $8$-vertex statistical model and similarly between
the $XXZ$ chain and the $6$-vertex model \c{baxter}. Hamiltonian $H_s$ of
the integrable quantum spin chain is given through its transfer
 matrix  as $ ln \tau(u)= I + u H_s + O(u^2)$, while the partition
function $Z$
 of the related vertex model is constructed from  $\tau(u)$ as $Z=
tr(\tau(u)^M)$. Moreover, both these models
 share  the same quantum $R$-matrix and  have  the same  
 representation for the transfer matrix,   
  commutativity of which:  $[\tau(u),\tau(v)]=0 $  
 guarantees their integrability.
 
It is therefore rather surprising to note that, inspite of such deep
connection between these two
 integrable systems,   their starting formulation conventionally follows two
different routes. Quantum systems  usually are  defined by  their
Lax operators
$L_{al}(u)$, which satisfy the quantum
Yang-Baxter equation (YBE) $ R_{ab}(u-v) L_{al}(u)
L_{bl}(v)=  L_{bl}(v)  L_{al}(u) R_{ab}(u-v),$ together  with its associated
 $R$-matrix. 
 A  vertex model on the other hand is described  by its Boltzmann weights
 given generally through the elements of the $R$-matrix
alone, which solves the YBE
$R_{ab}(u-v) R_{al}(u) R_{bl}(v)= R_{bl}(v) R_{al}(u) R_{ab}(u-v).$
 However such  a  difference in their approaches,  reason of which seems to
 be rather historical,
  puts certain restrictions on the 2-dimensional vertex models by
  assuming their vertical (v) and the horizontal (h) links,
 which are related   to the auxiliary and quantum spaces respectively,
 to be equivalent. As a consequence while 
 a rich variety of integrable quantum systems with wide range of
interactions involving   spin, fermionic,  bosonic and 
canonical variables does  exist,
    the  integrable vertex models are confined mostly to those 
quantum models that  exhibit  regularity
property expressed through the permutation operator: 
$L_{al}(0)=P_{al}$ and  hence   correspond 
to local Hamiltonians with nearest neighbor (NN) interactions. 
Therefore the well known examples of the integrable vertex models,
 apart from those mentioned above, appear to be   limited mainly to
the models like the 5-vertex model \c{5v},
  6-vertex model in external fields \c{g6v},
$19$-vertex model connected with the
 Babujian-Takhtajan spin-1 chain \c{babus} and  the   vertex models
  equivalent  to the Hubbard model \c{hubstat},
  supersymmetric t-J model \c{tjstat},
 Bariev chain \c{barstat} etc.  all
exhibiting only  NN interactions.

 The basic idea of the present letter however is to
exploit fully the equivalence between statistical and quantum systems
 and construct new class of integrable  vertex models by
applying a unifying  scheme  designed originally for 
  quantum  models \c{kunprl99}. In the original scheme an 
ancestor model was proposed for generating 
 integrable  quantum systems  as its various descendant
realisations. For describing our vertex models we start in
analogy also with the generalised Lax operator \c{kunprl99}
\be
L{(u)} = \left( \begin{array}{c}
  {c_1^+} q^{ S^3+u}+ {c_1^-}  q^{- (S^3+u)}\qquad \ \ 
  2\sin \al  S^-   \\
    \quad  
   2\sin \al S^+    \qquad \ \  {c_2^+}q^{- (S^3-u)}+ 
{c_2^-}q^{S^3-u}
          \end{array}   \right), \quad
         q=
e^{i \alpha } ,  \ll{L} \ee
 linked with
 the underlying quantum
algebra
\be
 [S^3,S^{\pm}] = \pm S^{\pm} , \ \ \ [ S^ {+}, S^{-} ] =
  ( M^+[2  S^3]_q + {M^- } 
[[ 2  S^3 ] ]_q) , \quad  [M^\pm, \cdot]=0.
\ll{Aalg} \ee
 Here $ [x]_q \equiv
{\sin (\al x)\over \sin \al},\ \ [[x]]_q \equiv
{\cos (\al x)\over \sin \al}
$ and the central elements
 $ M^\pm$ are related to the other set of  such elements appearing in the 
$L$-operator 
as
  $ M^\pm=\pm \ha  \sqrt {\pm 1} ( c^+_+c^-_- \pm
c^-_+c^+_- ). $ It is important to notice that
(\re{Aalg}) is a q-deformed quadratic algebra, which generalises both q-spin
and q-boson algebras and
 in fact  follows from the quantum YBE representing  integrability
condition. We define the Boltzmann weights (BW)
 of our vertex models 
 not by  the  $R$-matrix as conventional, but
 through the elements of the
   Lax operator: $L_{ab}^{j,k}(u) =\omega_{a,j;b,k}(u)$  by 
using 
matrix representations of the general algebra (\re{Aalg}) in (\re{L}).
  These generalised BW
 generate a unified  vertex model, which through 
possible reductions  yields   new series of  vertex models
 linked with
 different underlying algebras, their representations and choices of
the central elements.
 Prominent
examples of such  integrable statistical systems are  a 
 rich collection of   vertex models linked
to q-spin and q-boson with generic q, q roots of unity and $q
\to 1$.
 In all these models
 the h and v links
, contrary to the usual approach,
may  become  inequivalent and independent  at every vertex point  and
since we consider here  $2
\times 2$ Lax operators, the h links    admit
only 2 values: $a,b \in [+,-] $.
 The v links on the other hand  have
richer  possibilities with $j,k \in [1,D] $,
depending on dimension $D$ of 
the matrix-representation of the  q-algebras (see Fig. 1).
The familiar ice-rule is generalised here as the 'colour' conservation 
 $a+j=b+k$ for determining nonzero  BW.
 The crucial partition function of the models however is given as usual by 
 $Z= \sum_{config} \prod_{a,b,j,k} \omega_{a,j;b,k}(u).$
 
An important point to note is that
unlike  traditional approach the Lax operators related to  such vertex
models do not coincide with their
$R$-matrix, do not comply with the regularity condition and do not
correspond in general to quantum Hamiltonians
 with NN  interactions.
Moreover since our vertex models belonging to the same class have 
 the same $R$-matrix, 
we can generate another 
 rich series    of  integrable models, namely 
 hybrid vertex models  
by combining any number of  them in a row (see Fig. 1).
 
The eigenvalue solution of the transfer matrix needed for constructing the
partition function for all these vertex models can also  be found exactly 
through the algebraic Bethe ansatz in a unifying way.

\smallskip

\noindent {\it Unified vertex model}:

 In accordance with our primary
goal we discover an
  explicit matrix representation for the basic operators
$S^\pm,S^3$:
  \be
<s,\bar m|S^3|m,s>=m \delta_{m,\bar m},  \quad <s,\bar m|S^\pm|m,s>
= f^\pm_s(m)\delta_{m\pm 1,\bar m} ,\ll{qsrep}\ee 
 with $f^+_s(m)=f^-_s(m+1) =( \kappa+
[s- m]_q (M^+[s+ m +1]_q + M^-[[s+ m +1]]_q)) ^{\ha}
$ having 
 additional  parameters $\kappa, s$. It may be checked
that (\re{qsrep}) indeed gives an exact  representation of the general
q-deformed
 algebra (\re{Aalg})
  for  arbitrary values of the  central elements
  $M^\pm$. Therefore the   BW
may be constructed from the matrix representation  of the 
generalised  Lax operator  (\re{L})  by using  (\re{qsrep})  in the form
\be \omega_{\pm,j;\pm,j} (u)=
{c_\pm^+} e^{i \al (u\pm m)}+ {c_\pm^-} e^{-i \al (u\pm m)}, \  
 \ \ \ 
\omega _{+,j;- ,j-1}=\omega _{-,j-1;+ ,j}=  
2 f^+_s(m) \sin \al
 , \ll{Avertex}\ee
with $ \ m=s+1-j, \ \  j= 1,2,\ldots,D. $
  The  BW parameterised as (\re{Avertex}) would  generate now
 a unified $(4D-2)$-vertex model, representing  a new   series 
with arbitrary parameters
 $c^\pm_\pm,\ s$ and $\kappa$. These models and naturally all others
 obtained below through various reductions 
are   integrable statistical models and
 share the
same $R$-matrix, which is given by that of the well known
   $6$-vertex model \c{baxter}.

Note that though in general the dimension $D$ of the 
    matrices 
(\re{qsrep}) is infinite,  it  may get truncated through possible
appearance of  zero-normed states. To analyse this important effect
 we observe that 
since
$[0]_q=0$, one gets $f^+_s(m=s)=0$  for $ \kappa=0 $, recovering
 the familiar   'vacuum' state: $S^+|s,s>=0$.
However  due to the presence of the second term 
one gets here  $f^-_s(m=-s)=
( [2s+1]_q (M^+[0]_q + M^-[[0]]_q)) ^{\ha}
\neq 0$ and unlike the  spin representation we have
 $S^-|m,s> \neq 0$ for all $m$. This  creates therefore
  an infinite tower of  states by the action of the lowering operator $S^-,$
as typical for the bosonic representation.
  This also signals  the fact that algebra (\re{Aalg}) includes 
both  q-spin and  
  q-boson and therefore  their representations  must  be derivable 
from (\re{qsrep}) as particular cases.

\smallskip

\noindent  {\it q-spin vertex model}:

 It is straightforward to check that 
for $ M^+=1, M^-=0,$  our unifying algebra (\re{Aalg}) reduces to 
the well known $U_q(su(2))$ quantum spin algebra 
 \c{qa}
and at the same time  (\re{qsrep}) reproduces the known q-spin
representation. 
Therefore the corresponding BW may  be  
obtained  from   
(\re{Avertex}) for a consistent  choice  
 $c_+^\pm=c_-^\pm=\mp i,$  as $$
 \omega _{\pm,j;\pm,j}(u)=
[u\pm m]_q, 
\ \ \omega _{+,j;- ,j-1}=\omega _{-,j-1;+ ,j}=  
 f^{+(qspin)}_s(m), \ \ m=s+1-j $$ with 
$f^{\pm(qspin)}_s(m)=([s\mp m]_q [s \pm m +1]_q ) ^{\ha}$.
In this case the truncation $S^\pm|m=\pm s,s> = 0$ 
 typical for spin models and hence the familiar
 $D=2s+1$  dimensional representation
  naturally arise, which   produces therefore 
  a series of q-spin $( 8s
+2)$-vertex  models.
 The $6$-vertex model  is clearly  recovered  at $s= \ha$, while 
  $s=1, {3 \ov 2}, 2, {5 \ov 2}, \ldots$ 
  yield new $10, 14, 18, 22, \ldots$-vertex models (Fig. 1 a,b)).


 The
quantum systems related to such statistical models may be 
represented in general
 by  interacting q-spins with nonlocal interactions.
In particular, since the  well known sine-Gordon model is a realisation 
of the   q-spin \c{fadqs}, the vertex models constructed 
 with  nonzero $\kappa $ and having infinite $D$ will be related to the
 quantum integrable { lattice sine-Gordon model} \c{lsg}.
\smallskip

\noindent  {\it q-boson vertex model}:

 We find that q-bosonic algebra \c{qbos} 
  can also be derived as a subalgebra of
 (\re{Aalg}) for the complementary choice  $ M^+=\sin \al, M^-=\cos \al$
 by denoting 
$ S^+= \rho A, \ S^-= \rho A^\dag, \ S^3= -N, \ \ \rho
=( \cot \al)^\ha.$
   The same choice   derives therefore the matrix representation for the
q-boson directly from (\re{qsrep}) 
with the  assumption that $\kappa =s=0$ and   $n=-m$, yielding
$$f^{-(qbos)}_0(n)=([1+n]_q[[-n-1]]_q)^\ha= {1 \ov \sqrt 2}[1+n]^\ha_{q^2}, \ 
f^{+(qbos)}_0(n)=f^{-(qbos)}_0(n-1)={1 \ov \sqrt 2}[n]_{q^2}^\ha . $$ 
  Consequently for a consistent solution $c^+_\pm=1, c^-_\pm=\mp i e^{\pm
i \al}$  we can   
derive from (\re{Avertex}) the  
BW 
as
 \bea \omega_{\pm,j;\pm,j}(u)&=&
 ie^{\pm i \al  \phi} 
[u \mp (j+\phi -1)]_q ,\ \ \ \phi = \ha (1+ {\pi \ov 2 \al}), \nonumber \\
\omega _{+,j;- ,j-1}&=&\omega _{-,j-1;+ ,j}=
 f^{+(qbos)}_0(j-1)=
 {1 \ov \sqrt 2} [j-1]^\ha_{q^2} .  
\ll{qbboltzman}\eea
 

It is obvious that apart from  the vacuum state $|0>$ with 
$f^{+(qbos)}_0(0)={1 \ov \sqrt 2} [0]^\ha_{q^2}=0$
 we can have no  other zero-normed states and the
  q-bosonic representation like  the standard boson
 is semi-infinite with $D=n+1$.
    The integrable
 $(4n+2)$-vertex
model linked to the q-boson (Fig. 1c))
 that we construct using (\re{qbboltzman}) would therefore
 be related to
the    lattice  version of the quantum { derivative nonlinear Schr\"odinger 
model} (DNLS), which exhibits a  q-bosonic realisation \c{qdnls}. 

\medskip

\noindent {\it Vertex models with $q$ roots of unity}:

An excellent possibility for regulating the dimension  of the matrix
representation opens up when $q=e^{i \al}$ is chosen as  solutions of
$q^p=\pm 1$
with parameter $\al$ taking discrete values $\al _a = \pi {a \ov p}, a= 1,2,
\ldots,p-1$   \c{saclay}. Note however that when some values of $a$ becomes 
a factor of $p$ one faces a situation with  $q^{p \ov a}= \pm 1.$ Therefore
to avoid such complicacies we suppose $p$ to be prime in our present
 discussion. 
For further analysis  we focus on the action of $S^-$ assuming
$\kappa=0$ 
 in  (\re{qsrep}):
$ S^-|m=-\bar s,s> =
( [s+\bar s +1 ]_q (M^+[s-\bar s ]_q + M^-[[s-\bar  s]]_q)) ^{\ha}.
$
and  observe  that
   due to $[p]_q=\sin \al_a p= 0,$  unlike generic q we can get now  
$S^-|-\bar s,s> = 0$ 
 at $ \bar s= p-(s+1)$,  which
  reduces   
matrix (\re{qsrep}) to  a  finite  dimensional  representation.
  Therefore the BW obtained from (\re{Avertex})
for this case would
 generate another
  series of unified
 ${\sc K}$-vertex  model
 having finite
  ${\sc K}= 4p-2$ configurations
 at every vertex point. 
Moreover, since for a fixed  p there can be 
  $p-1$ different   
  $\al_a$,
 each of these discrete values   describes a different set of BW and hence
 a novel  model.

  Consequently
  at  particular reductions as analysed  above,   we  obtain 
  the corresponding  series of new  vertex
models linked with q-spin or q-boson, but now having  finite configuration
space determined  by  $p$.
Thus for the 
 q-spin with fixed $p, 0<p<2s+1,$
 in place of a  $(8s+2)$-vertex model for 
  generic $q$, one obtains  $p-1$ number of different 
 $ (4p-2)$-vertex models and the related representation
 including the case 
$p>2s+1$ become   more involved 
\c{saclay}. The corresponding BW 
  defining these models should  however  be given by their same
 generic form, though 
using discrete $\al_a$ values.
   As for
example in case of $s={5 \ov 2} $ with
$q^5=- 1$,      instead of a $22$-vertex model 
      one obtains $4$ 
 different $18$-vertex models for 
    distinct values of $\al_a=\pi {a \ov 5}, a=1,2,3,4$.
Noticeably, as a quantum
model the q-spin with q roots of unity  are realised as
  the { restricted  sine-Gordon model} \c{rsg}.
 
The situation  becomes  more interesting when applied to the  
q-boson   with finite $p$,
 since  now together  with the standard vacuum
 we  get also   $A^\dag |n={p \ov 2}-1>
= {1 \ov \sqrt 2}[{p\ov 2}]^\ha_{q^2}|{p \ov 2}>=0 $, yielding 
finite $({p \ov 2} \times {p \ov 2})$
  matrix representations for the q-bosonic operators $ A, A^\dag $.
As a result we obtain an intriguing series of $(2p-2)$-vertex models with
 BW described by the same form (\re{qbboltzman}) as for the generic
q-bosonic case, but with
different possible  parameter values
$q=e^{i \al_a }, a=1, 2, \ldots, p-1$.
The  quantum model corresponding  to 
such  q-boson vertex models can be realised as the 
  { restricted 
 DNLS} model, which  supports  finite quasi-particle 
 bound states \c{bt01}.

\noindent {\it Rational class of vertex models }:

At  $q \to 1 ( \al \to 0 )$ on the other hand,
 the associated $R$-matrix
goes to its known  rational limit 
 and the underlying algebra becomes  undeformed one 
 with $M^\pm \to m^\pm $, reducing at the same time  
the unified  model 
to its rational form.
 Consequently, taking carefully the limits we may construct  in a similar way 
 the corresponding set of vertex models belonging to the rational class. Not
going into details we mention only that the BW of these
vertex models can be obtained from  the limiting values of (\re{Avertex})
yielding
$ f^+_s(m)\to ( (s-m)(m^+ (m+s+1)+m^-))^\ha.$
 It is easy to check that 
the BW for the vertex models related to 
the undeformed spin as well as the standard 
boson  correspond to the  particular values of the central elements:
 $m^+=1, m^-=0 $ and $ m^+=0, 
m^- =1$, respectively. Remarkably,
   the spin vertex model  constructed in this way  coincides 
with the similar higher $s$ model obtained earlier
through fusion method \c{babus}, whereas    
 the bosonic-vertex  model 
 apparently is a new model, linked to a quantum integrable
{ lattice  NLS} model \c{kunrag}.

\noindent {\it Hybrid vertex models}:

In  constructing our vertex models  we have flatly assumed  that in any
model the same BW must be defined at every vertex point. An immediate
generalisation is therefore possible by relaxing this condition and
considering the central elements
 $c^\pm_\pm$ as well as the  spin  parameters $s$ appearing in 
(\re{Avertex}) to be different at different sites.
As we have already stressed,   vertex
models obtained as various reductions
 of the same  integrable unified model  belong to the same class
 sharing  the same $R$-matrix. Thus the 
q-spin and q-boson vertex models are members of the trigonometric class,
while the normal spin and boson models belong to the rational 
class. Based on this fact  therefore we 
  can construct a rich collection of { hybrid models} by combining different
 vertex models of the same class and inserting their defining BW
along the vertex points $l=1,2,\ldots,N$ in a row,
 in any but fixed manner. 
Due to the  association 
with the same $R$-matrix the integrability of such statistical
models would be naturally preserved.

Thus for example an alternate insertion of  $10$ and $6$ vertex models
 results to a hybrid model, which is related to the known quantum
model
\c{alts}
involving  spin-1 and spin-$\ha$ operators with next-NN interactions. 
 More exotic hybrid  models can  be formed
by arranging the BW for the q-spin and q-boson
vertex models,
 alternatively or in any other way at different   vertex points (Fig. 1). 
Similarly one can construct a 
 spin-boson hybrid vertex model by
combining their individual  vertex models, which would  correspond 
 to a quantum chain of interacting  spins and bosons
 involving next-NN couplings.

\medskip

\noindent {\it Unified  solution}:

The construction  of the  unified  vertex model 
 through the generalised Lax operator suggests also 
a   scheme for  exactly solving the eigenvalue problem 
for the transfer matrix. Since the partition functions in turn
can be   determined from the knowledge of these eigenvalues, all vertex  models
obtained  as  particular cases and  linked to (un-)deformed  spin  or (un-)deformed  boson   
 can also be solved  in a unified way.
There is a well formulated algebraic Bethe ansatz method
for exactly solving the 
  eigenvalue problem of the transfer matrix: 
$\tau(u)=tr_h (\prod_l^N L_l(u))$, 
when the associated  Lax operator  
     as well as  the  $R$-matrix 
are given \c{aba}. Therefore, 
since we have defined the BW through matrix representations
 of the  Lax operator and the $R$-matrix in our case is given by that of the
 well known $6$-vertex model, we can derive the exact eigenvalues for the
transfer matrix of our models as
\be
{\Lambda}(u)=  
\omega_{+,1;+,1}^N(u)\prod_k^n {g}(u_k-u)+
\omega_{-,1;-,1}^N(u)\prod_k^n {g}(u-u_k)
, \ \ g(u)= { [u+ 1]_q \ov [u]_q}. 
\ll{ev}\ee
with all possible solutions of $\{u_k\}$ to be  determined from 
 the Bethe equations \be
\left ({\omega_{+,1;+,1}(u_l)\over \omega_{-,1;-,1}(u_l)}\right )^N=
\prod_{k \neq l}^n {[u_l-u_k+1]_q \over [u_l-u_k-1]_q }, \ \  \ l=
1,2,\ldots, n.
\ll{be}\ee 
By analysing the structure of these equations  
 we  conclude that,   the  factors involving BW in both of them come from
the action of the Lax operator on the pseudovacuum, which is chosen as the
direct product of the highest weight states with $j=1$ i.e. $|m=s>$. The
rest of the factors  on the other hand
 are  originated from the $R$-matrix elements, which  arise during
diagonalisation of the transfer matrix due to the use of the quantum YBA. 
Therefore it is crucial to note that, the only 
 part  given
by  BW is model-dependent and defined for the  vertex models
 by the diagonal entries in (\re{Avertex}) with $j=1$, while
  the remaining parts contributed by  the $R$-matrix are
    the same for all our models from the same  class.  Consequently the exact
solutions for all  models constructed here 
 can be found in a systematic way from
 (\re{ev}) and (\re{be}) by using corresponding reductions 
of the  unified model (\re{Avertex}).

The
total number of  solutions $\{ \Lambda_\gamma (u)\} $ 
for  the eigenvalues 
(\re{ev}) should be $D^N$, which 
coincides with the number of possible eigenstates and gives
the dimension of the vector space on which the transfer matrix  
acts. The partition  function of the vertex models may therefore be 
 given at the thermodynamic limit by
  $Z=lim_{M,N \to \infty}tr_v(\tau^M(u))=
lim_{M,N \to \infty}\sum_{\gamma=1}^{D^N} \Lambda_\gamma^M(u)$.
At this important   limit,
  the Bethe equations (\re{be}) turn
  into an integral equation  
 $ V(u)= 2 \pi \rho(u)+\int \rho (v)K(v,u) dv $, with  known kernel
of the 6-vertex model \c{6v}. Interestingly,
 all information about a  particular 
model is 
encoded in the driving term only, which is expressed through 
${\omega_{+,1;+,1}(u) \ov \omega_{-,1;-,1}(u)}
=r e^{i P(u)}$ as $V(u)= P^{'}(u)$ and 
 therefore  knowing the explicit form of  BM one can   
 derive easily  the equations for  individual models.

For extracting the solutions of the hybrid vertex
models however the BW dependent parts in the above equations should be
 slightly modified by generalising the factors inhomogeneously as
$\prod_\beta ((\omega_{\pm,1;\pm,1}^{(\beta)}(u))^{N_\beta} $, where 
$N_\beta$ is the number of vertices of type $\beta$ appearing in a row
with the constraint $N=\sum_\beta N_\beta$.

Detail investigation of individual  models and identification of their 
most probable states are   
 important problems  to be pursued.


\newpage
\begin{figure}[ht]
\begin{center}
\leavevmode
\epsfxsize=0.55\textwidth
\epsfbox{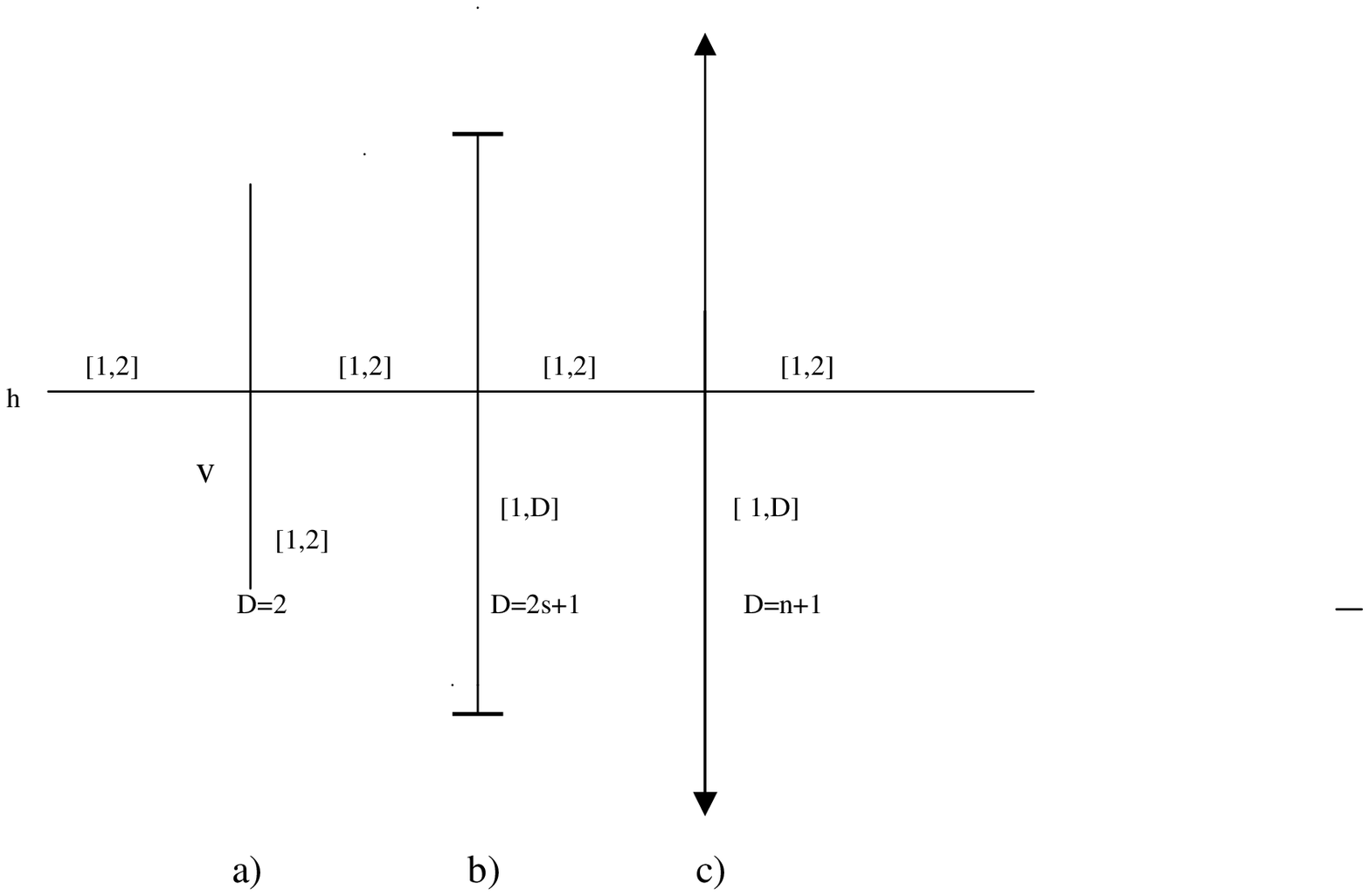}
\end{center}
\caption{
   Integrable vertex models with horizontal (h)
 links taking $2$ values,
while the vertical (v) ones may  have $D$ possible values.  
a) 6-vertex   b) q-spin vertex  and  c) q-boson vertex models. 
Combining a,b,c) an integrable hybrid model may be formed.  $q^p=1$ gives
 $D=p$ in b) and c)} .
\label{fig:phases}
\end{figure}


\begin{thebibliography}{99}
\bibitem{baxter} R. Baxter, { Exactly Solved Models in Statistical
Mechanics} (NY, Academic, 1982)
\bibitem{5v} F. Y. Wu, Phys. Rev. {\bf 168 } 
(1968) 539
\bibitem{g6v} B. Sutherland, C P Yang and C N Yang, Phys. Rev. Lett. {\bf
19}
(1967) 588.
\bibitem{babus} J. Babujian,  Phys. Lett {\bf 90 A}  479  (1982)
\bibitem{hubstat} E. Olmedilla, M. Wadati and Y. Akutsu, J. Phys. Soc. Jpn. 
 {\bf 87}  2298 (1987)
\bibitem{tjstat} F. Essler and V. Korepin, Phys. Rev. {\bf B 46 } 9147 
(1992) ;  A. Foerster and M. Karowski, 
 Phys. Rev.  {\bf B 46}  9234 (1992)
\bibitem{barstat}  H. Q. Zhou, J. Phys. {\bf 29}  5504 (1996)
\bibitem{kunprl99}
Anjan Kundu, Phys. Rev. Lett. {\bf 82} 3936  (1999) 
\bibitem{qa} V. G. Drinfeld, Quantum Groups (ICM Proceedingd, Berkley,
1986) 798; 
M. Jimbo, Comm. Math. Phys. {\bf 102}
  537 (1987);
N. Reshetikhin, L. Takhtajan and L. Faddeev, Algebra and Analysis {\bf 1}
178 (1989)
\bibitem{fadqs} L.  Faddeev, Int. J. Mod. Phys. {\bf A 10} 1845  (1995)  
\bibitem{lsg} A.  Izergin  and V.  Korepin, 
{ Nucl. Phys.} {\bf B 205} [FS 5]  401  (1982)
\bibitem{qbos} A. J. Macfarlane,  J. Phys. {\bf A 22}   4581 (1989);
L. C. Biederharn,     J. Phys. {\bf A 22}   L873  (1989); C. P. Sun and H.
C. Fu,   J. Phys. {\bf A 22}   L983  (1989)
\bibitem{qdnls} 
 Anjan Kundu  and B. Basu-Mallick, J. Math. Phys. {\bf  34}  1252  (1993) 
\bibitem{saclay} 
V. Pasquier  and H. Saleur,  Nucl. Phys.  {\bf B 330}  523  (1990) 
\bibitem{rsg} A. Leclair, Phys. Lett.  {\bf B 230}  103  (1989)
\bibitem{bt01} B. Basu-Mallick snd T. Bhattacharyya,
  Nucl. Phys.  {\bf B 634}  611 (2002)
 {\it ABA for a quantum
integrable DNLS}, Preprint: arXiv:hep-th/0202035 (2002) (to appear in
Nucl.Phys. {\bf B}
\bibitem{kunrag} Anjan  Kundu and O. Ragnisco, J.  Phys. {\bf A 27}
   6335 (1994)
\bibitem{alts}  H. de Vega and F. Woynarovich, J. Phys. {\bf A 25} 4499 (1992) 
\bibitem{aba}  
 L. D. Faddeev, Sov. Sc. Rev. {\bf C1} 107 (1980)
\bibitem{6v} C. N. Yang and C. P. Yang  Phys. Rev. {\bf 150} 321, 327
 (1966) 


\end{thebibliography}
 \end{document}